\documentclass{article}

\usepackage[round]{natbib}

\usepackage{enumitem}
\usepackage[centertags]{amsmath}
\usepackage{amsfonts,amsthm,amssymb}
\usepackage{amssymb}
\usepackage{amsmath}
\usepackage{url}
\usepackage{soul}
\usepackage{graphicx}
\usepackage{accents}
\usepackage{booktabs}
\usepackage{appendix}
\usepackage[hmargin=3.175cm,vmargin=3.175cm]{geometry}
\usepackage{etoolbox}
\usepackage{xparse}
\usepackage{enumitem}

\usepackage{color}
\usepackage[svgnames]{xcolor}
\usepackage{soul}
\usepackage{chngcntr}
\usepackage{authblk}

\newcommand{\ubar}[1]{\underaccent{\bar}{#1}}
\DeclareDocumentCommand{\publicBelief}{O{\mu}}{#1}
\DeclareDocumentCommand{\privateBelief}{O{p}}{#1}
\DeclareDocumentCommand{\signalupdate}{O{q}}{#1}
\DeclareDocumentCommand{\signal}{O{s}}{#1}
\DeclareDocumentCommand{\signalsSet}{O{S}}{#1}
\DeclareDocumentCommand{\state}{O{\omega}}{#1}
\DeclareDocumentCommand{\statesSet}{O{\Omega}}{#1}
\DeclareDocumentCommand{\price}{O{\tau}}{#1}
\DeclareDocumentCommand{\action}{O{a}}{#1}
\DeclareDocumentCommand{\limitParam}{O{\alpha}}{#1}
\DeclareDocumentCommand{\deterrencePrice}{O{\price} O{d}}{#1^#2}
\DeclareDocumentCommand{\LLR}{O{x}}{\log(\frac{#1}{1-#1})}
\DeclareDocumentCommand{\lBound}{O{\limitParam} O{\publicBelief}}{\ubar{#1}_{#2}}
\DeclareDocumentCommand{\uBound}{O{\limitParam} O{\publicBelief}}{\bar{#1}_{#2}}
\DeclareDocumentCommand{\eqPrice}{O{\price}}{#1^{*}}

\newtheorem{lemma}{Lemma}

\newtheorem{observation}{Observation}
\newtheorem{theorem}{Theorem}

\newtheorem{corollary}{Corollary}

\theoremstyle{definition}
\newtheorem{definition}{Definition}

\newlist{secenum}{enumerate}{10}
\setlist[secenum]{label=\thesection.\arabic*,leftmargin=*}

\newtheorem{innercustomgeneric}{\customgenericname}
\providecommand{\customgenericname}{}
\newcommand{\newcustomtheorem}[2]{%
	\newenvironment{#1}[1]
	{%
		\renewcommand\customgenericname{#2}%
		\renewcommand\theinnercustomgeneric{##1}%
		\innercustomgeneric
	}
	{\endinnercustomgeneric}
}

\newcustomtheorem{customthm}{Theorem}
\newcustomtheorem{customlemma}{Lemma}
\newcommand{\blocktheorem}[1]{%
	\csletcs{old#1}{#1}
	\csletcs{endold#1}{end#1}
	\RenewDocumentEnvironment{#1}{o}
	{\par\addvspace{1.5ex}
		\noindent\begin{minipage}{\textwidth}
			\IfNoValueTF{##1}
			{\csuse{old#1}}
			{\csuse{old#1}[##1]}}
		{\csuse{endold#1}
		\end{minipage}
		\par\addvspace{1.5ex}}
}

\blocktheorem{customlemma}

\usepackage{array}
\usepackage{makecell}

	\definecolor{ao}{rgb}{0.0, 0.5, 0.0}

\usepackage{hyperref}
\usepackage{subcaption}
\usepackage{multirow}

\usepackage{colortbl}

\usepackage{caption}

\usepackage{amsmath}
\usepackage{amssymb}

\title{The Multi-BMBY Mechanism: Proportionality-Preserving and Strategyproof Ownership Restructuring in Private Companies}
\author[1]{Gal Danino\thanks{gald@technion.ac.il, korenmor@bgu.ac.il and omermadmon@campus.technion.ac.il. Equal author contribution.}}
\author[2]{Moran Koren}
\author[1]{Omer Madmon}

\affil[1]{Technion - Israel Institute of Technology, Haifa, Israel}
\affil[2]{Ben Gurion University of the Negev, Beer Sheva, Israel}

\renewcommand\footnotemark{}

\date{\today}

\begin{document}

\maketitle
\begin{abstract}
    In privately held startups, restructuring ownership is challenging due to diverse and uncertain valuations among owners. Traditional approaches, including the BMBY mechanism for equal partnerships, fail to address the complexities of multi-owner settings and don't elicit true valuations. We propose a novel mechanism that extends the BMBY rationale to accommodate these complex scenarios. Our mechanism ensures truthful valuation elicitation while offering several advantages: it is easy to implement, budget balanced, resistant to collusion, individually rational, and allocates shares to those who value them most. Crucially, it preserves proportionality among remaining owners, maintaining existing power dynamics. The mechanism allows for adaptive control of the eventual number of owners, addressing unique startup needs such as incentivizing employee ownership. This paper contributes to the field of ownership restructuring by providing a practical, theoretically-grounded solution for the complex dynamics of startup recapitalization, potentially improving decision-making processes and stakeholder relationships in these pivotal business transitions.
\end{abstract}

\noindent \textbf{Keywords:} game theory, mechanism design, fair allocation, ownership restructuring, partnership dissolvement, share buybacks

\noindent \textbf{JEL Classification:} C72, D82

\newpage
\section{Introduction}
\newcommand{\note}[1]{\textcolor{red}{#1}}


In the dynamic world of privately held startups, ownership structures often become complex and misaligned over time. Consider a technology startup that has pivoted several times since its inception. Over the years, it has accumulated a diverse array of owners—early investors, venture capital firms, angel investors, and former employees who received equity compensation. While this broad ownership base was crucial for the company's growth, it now presents challenges as the company enters a new strategic phase (e.g., prepares for a potential acquisition or pivot). 
The current owners wish to streamline the cap table, reducing the number of shareholders to simplify negotiations and decision-making processes. This practice is often referred to as \emph{recapitalization}, and is widely accepted as a best practice in startup management \citep{moro2021startup}.

However, the shareholders face a delicate balancing act: they need to preserve the existing power dynamics among the remaining owners while ensuring that those who value the company most retain their stakes.
Furthermore, maintaining a significant number of owners post-restructuring is vital, as ownership tends to enhance job quality and boost employee commitment (see e.g. \citealp{gornall2024employees}).
Traditional methods fall short in addressing these competing objectives simultaneously. 

Share buybacks, a common approach for public companies, face a fundamental challenge in private firms: the lack of a predetermined market price. This absence makes it difficult to set a fair buyback price that reflects the true value of the company and accounts for the differing valuations of diverse owners.\footnote{
The "buyback" mechanism in public firms exists due to the availability of a market price which serves as an anchor to the seller compensation. This anchor price does not exist in privately held firms which constitute the vast majority of companies in the US (e.g., in 2013, less than 15\% of the firms with over 500 employees were public and in 2010 private firms accounted for 59\% of US sales, \citealp{Biery2019}). For this sector, which comprises various firm sizes ranging from small mom-and-pop shops to huge conglomerates (\citealp{Forbes2019}), the buyback mechanism cannot simply be applied and often requires a costly due diligence process to assess the company share value. 
} An alternative approach is sequential negotiation, which can be time-consuming, prone to strategic manipulation, and may not result in an efficient allocation. Both methods struggle to preserve proportionality among remaining owners while ensuring that those with the highest valuations retain their stakes. 

This scenario illustrates a common challenge in ownership restructuring of privately held assets—how to efficiently reallocate shares while maintaining proportionality among remaining owners and maximizing overall welfare, all without the benefit of a market-determined price. In this paper, we present a novel mechanism that tackles these challenges head-on.
	
When the firm is co-owned by two partners, each owning half, there exists an ownership restructuring mechanism called "BMBY - buy me out or I will buy you out". The BMBY mechanism works as follows, say Alice and Bob co-own an indivisible asset, and wish to part ways. In BMBY, one partner, say Alice, declares a price and the other partner, Bob, decides whether he wants to buy Alice out or sell his share to Alice at the aforementioned price. This simple mechanism is extremely useful and intuitive, yet may be susceptible to strategic behavior and may lead to an inefficient allocation.\footnote{To see why, assume that agent types are determined by i.i.d draws from a uniform distribution over $[0,1],$ and consider a case where Alice's valuation is extremely low. In this case, as Bob's expected value is half, she can benefit from misreporting her value and declaring a higher price.
	}

\subsection{Our contribution}
In this work, we extend the simple two-agent BMBY mechanism to a setting in which the asset is owned by many share holders, while maintaining the mechanism simplicity.  We introduce a mechanism for decreasing the number of owners of a stock from $n$ to at most $\bar{m}<n$ (where $\bar{m}$ is exogenous), which we call Multi-BMBY Mechanism (MBM). The mechanism receives bids from all agents, and sets an asset price, at which agents trade shares of the asset. The eventual number of owners $m$ is a random variable that can be either $\bar{m}$ or $\bar{m}-1$, with a carefully chosen probability. 
 The mechanism is strategyproof: for each agent either increasing or decreasing her truthful bid does not increase her gain, regardless of the bidding of other agents. Additionally, the mechanism is immune to collusions (weak group strategyproof), budget balanced, and individually rational.

Regarding efficiency, it is evident that the most effective allocation method is to assign the entire asset to the agent who values it the most. However, this first-best allocation cannot be achieved by any incentive-compatible and budget-balanced mechanism, even in the bilateral trading case, as shown by the celebrated paper of \cite{myerson1983efficient}. While previous work has focused on developing such mechanisms that are \emph{approximately} efficient (\citealp{Reallocation-mechanisms,blumrosen2021almost}), our work considers the case in which a more distributed allocation is desired due to \emph{normative} reasons, such as safeguarding the rights of minority shareholders or preserving employees' stakes in ownership. 

We therefore introduce a mechanism that receives $\bar{m}$ as an input, and outputs an allocation in which the proportionality of the remaining ($\bar{m}$ or $\bar{m}-1$) owners' shares is preserved with respect to their initial shares. If an allocation satisfies this condition and allocates the asset to the set of agents with the highest valuations, we say that the allocation satisfies \emph{proportionally-preserving ex-post efficiency}. Our proposed mechanism indeed satisfies this property.

Our approach differs from traditional ownership restructuring methods in two fundamental ways. First, instead of aiming for complete dissolution, we allow the final number of owners to be determined exogenously (up to one shareholder, in order to maintain strategyproofness). Second, we ensure that the allocation preserves the proportional shares of the original allocation. These constraints, while prioritizing proportionality and adaptability, may come at the cost of overall social welfare—typically the main focus of standard approaches for restructuring mechanisms.\footnote{Trivially, as the number of owners decreases, the output allocations approach the first-best allocation, as only a small number of high-valued owners remain. Therefore the requirement of fixing the eventual number of owners at the outset generally harms social welfare. However, as highlighted before, there is a variety of reasons to ensure a certain level of post-restructuring ownership. Welfare implications of our approach are discussed in Subsection \ref{subsec: welfare}.}
Imposing those axioms reflects some level of mismatch between the agents' welfare and the designer's objective, as is often the case in the startup recapitalization scenario. 

\section{Related Literature}

In this work we study how to perform "buyback" in privately held firms. This can be thought of as a case of a multilayer partnership dissolvement problem which was introduced by \cite{Cramton1987}. \cite{Cramton1987} study whether  a partnership of multiple parties can be dissolved efficiently. As in our case, the partner valuations are held privately and the mechanism must elicit truthful behavior in order to produce an efficient allocation.  In their model, inspired by the FCC's cell phone licensing allocation, they attempt to allocate a co-owned, indivisible asset to a single owner which values it most. They conclude that such a mechanism exists as long as no owner possesses a too large share in the original partnership. Additionally, they propose a bidding mechanism that generates the ex-post efficient allocation. Their mechanism is a sealed bid auction in which agents are either paid or participate in an all-pay auction. This bidding mechanism, is shown to be incentive compatible, individually rational, and ex-post efficient, yet it is not budget balanced. Recent work following \cite{Cramton1987} has examined the problem of partnership dissolvent to account for complementaries (see for example \citealp{Ensthaler2014a,Landeo2013b}). Our work is closest to \cite{Wasser2013a}, who introduced a $k+1$ auction where the price is a convex combination of the bidder's bids, and thus not incentive compatible. Our proposed mechanism guarantees budget balanceness. In addition, it generates an equal share price for all participants, which is crucial for the case of equity restructuring. Additional work on partnership dissolvement examined the legal and social aspects of the problem, e.g. \cite{Adler2001} and \cite{Bredin2005}.

Our work strongly relates to \cite{myerson1983efficient}, who studied mechanism design in the context of bilateral trading with private information. \cite{myerson1983efficient} showed that there exists no mechanism that is strategyproof, budget balanced, individually rational and ex-post efficient. This model was then extended by \cite{loertscher2023bilateral} to a model of multi-unit with supply and demand, in which the agents' utility is geometric, meaning it can represent either increasing or decreasing marginal cost.

We show that under a less conservative notion of ex-post efficiency, these properties can be satisfied in a more general setting of ownership restructuring. In this setting, all agents are initially considered owners of the asset, and the mechanism then determines which agents are buyers (and remain owners) and which are sellers (and sell their shares). 
Here we allow for mechanisms that randomly draw the number of owners after restructuring. 
\footnote{In practice, our mechanism is non-deterministic only with respect to a single agent, meaning that the number of owners after the restructuring is approximately deterministic.}

Closer to our work, \cite{loertscher2020dominant} study a strategyproof, trade-sacrifice mechanism in a model in which all agents have maximum demands that are less than the aggregate supply. Their mechanism works by identifying a threshold agent that separates the buyers and sellers. In this work, we rely on a similar technique to derive a strategyproof, trade-sacrifice mechanism in a different model, in which each agent has a maximum demand for the entire supply.

An additional related line of work is the study of \textit{double auctions}. In a double auction a set of buyers and a set of sellers are interested in performing trades with each other. \cite{McAfee1992}, in his seminal paper have suggested a mechanism ,similar to ours, in which the price facing the sellers will be set by the declarations of the buyers and vice versa. The number of trades will be determined by supply and demand. A similar approach was applied to matching markets by \cite{Azevedo2016}. \cite{Dutting2017} extended the modular approach, which is often used for one sided matching problems, into the realm of double auctions, and identify the conditions under which incentive compatibility and feasibility are guaranteed. \cite{Bredin2005} examined the ability of a dynamic pricing policy to reduce the volatility and improve efficiency in online double auction markets. Our model covey an intuition similar to the McAfee auction, only we extend the analysis to a case where agents can be both buyers or sellers.

Finally, the problem of partnership dissolvement can be thought of as a cake cutting problem, where the size of the ``cake" is the value of the co-owned asset plus the sum of agent payments. In this set of models, the lion share of existing literature develop algorithms for dividing an asset among its owners. The main emphasis is on algorithms which are both  ``fair"   and efficient (see \citealp{Procaccia2013} for a recent review). The existing solutions in this field rely heavily on the order in which the agents are approached, and thus can be immensely time consuming to apply. Our mechanism is invariant to the order as bids are assumed to be simultaneous.
\section{The Model}
Let $I = \{1, ..., n\}$ be the set of agents such that $n > 2$. For any subset of agents $J \subseteq I$, we denote by $|J|$ the number of agents in $J$. Each agent $i \in I$ has a private value $v_i \in \mathbb{R_+}$ for the asset.
Let allocation space $\mathcal{A} = \Delta^n \times \mathbb{R}^n$ consist of allocations for shares and money for all $n$ agents, where $\Delta^n$ is the $n-$dimensional simplex.
Denote respectively by $s_i^A, x_i^A$ the share and money of agent $i$ under allocation $A$.

The MBM is a mapping $F: \mathcal{A} \times \mathbb{R}^n \rightarrow \Delta(\mathcal{A})$ from an initial allocation and a bid vector to a distribution over two new allocations: one of them with $\bar{m}$ owners, and the other one with $\bar{m}-1$ owners.
For each agent $i$ and for each new allocation, the mechanism defines the agent's monetary payment, which can be positive or negative, and the resulting share, which is non-negative. 
	
For any allocation $A \in \mathcal{A}$, Denote by $u_i (A) = s^A_i v_i + x^A_i$ the utility of agent $i$  from shares $s^A_i$, and money $x^A_i$. We normalize the initial amounts of money to be zero, and therefore the initial utility of each agent is simply $s^{A_0}_i v_i$, where $A_0$ denotes the initial allocation.
We assume agents face no monetary constraints.

The planner employs the mechanism to reallocate shares and money among the owners in a way that a subset $M \subset I$ of the initial owners remain owners at the end of the process. For the reallocation process, the mechanism receives declarations from all $n$ agents of their valuation of the asset as bids $b=(b_i)_{i \in I}$  and selects $1 \le m < n$ agents who remain owners.  The other $n-m$ initial owners sell their holdings.
The mechanism also decides on a single asset price $p$ that is used for all sales and purchases of shares.
Denote by $\bar{b}=(\bar{b}_{i})_{i \in I} = (v_{i})_{i \in I}$ the truthful bidding profile, and by $F(A_0, b | m)$ the allocation with $m \in \{ \bar{m}, \bar{m}-1 \}$ owners in the support of the $MBM$, under the bidding profile $b=(b_j)_{j \in I}$ and the initial allocation $A_0$.
When agents bid truthfully, i.e. $b = \bar{b}$, then we denote $\bar{A}_m = F(A_0, \bar{b} | m)$.
	
The mechanism works as follows: the planner limits the number of eventual owners with $\bar{m}$ that satisfies $1 < \bar{m} < n$.
 Each agent bids $b_i$ as a price for the whole asset, which represents her proportional declared valuation to every share size. Agents are ranked by their bids in \textbf{descending} order. We assume there are no ties in the agents' valuations.\footnote{Alternatively, one can assume that the agent valuations are drawn i.i.d. from an atom-less distribution and that ties occur with probability zero.} 
Let subscript $(i)$ denote the order statistic of bids such that $s^{A}_{(1)}$ is the share of the \textbf{highest} bidder. 
Let $r(b,i)$ denote the ranking of agent $i$ in the bidding profile $b=(b_j)_{j \in I}$ (i.e., agent $i$ for which $r(i)=1$ is the agents with the highest bid). When clear from context, we suppress $b$ and write $r(i)$ for brevity.
Then, the mechanism chooses the number of eventual owners $m\in\{\bar{m} - 1,\bar{m}\}$ at random (we describe the probabilities below), where $m = \bar{m}$ means that the $\bar{m}$'th ranked agent becomes a buyer, while $m = \bar{m} - 1$ means that the $\bar{m}$'th ranked agent becomes a seller. 
	
We set the probability of $m = \bar{m}$ and $m = \bar{m} - 1$ in a way that makes the $m$'th ranked agent receive a utility of zero in expectation.  
As we show later, the resulting probability for $m = \bar{m}$ is $\sum_{i=1}^{\bar{m}} s^{A_0}_{(i)}$ , (sum of the initial shares of the $\bar{m}$ highest bidders) and the probability for $m = \bar{m} - 1$ is $\sum_{i=\bar{m}+1}^{n} s^{A_0}_{(i)}$. 
	
The mechanism then sets $p=b_{(\bar{m})},$ i.e., the price equals the $\bar{m}$'th highest ranked agent's bid. A higher bidder agent, say agent $j$, becomes a buyer. Her share in the final allocation increases by $s^{A_0}_{j}\frac{\sum_{i=m+1}^{n} s^{A_0}_{(i)}}{\sum_{i=1}^{m} s^{A_0}_{(i)}} > 0$ (i.e., by a factor of $\frac{1}{\sum_{i=1}^{m} s^{A_0}_{(i)}} > 1$), and all lower bidder agents become sellers to whom the mechanism allocates no shares. All low bidders receive compensation, which equals the price times their respective share in the initial allocation. 
Denote by $\bar{r}, \bar{p}$ the ranking and price, respectively, when all agents bid their private valuation. 
	
	
\subsection{Definitions}
We begin by providing a formal definition of the MBM:

\begin{definition}
        The Multi-BMBY Mechanism (MBM) $F$ receives as an input the initial allocation $A_0$ and a bidding profile $b=(b_i)_{i \in I}$, and outputs a distribution over two allocations: $F(A_0,b|\bar{m})$ with probability $P(m=\bar{m})=\sum_{i=1}^{\bar{m}} s^{A_0}_{(i)}$ and $F(A_0,b|\bar{m}-1)$ otherwise.
        For both allocations, the asset price is set to be the $\bar{m}$'th highest bid, $p = b_{(\bar{m})}$.
        Each allocation is then defined as follows:

        $$s_{(j)}^{F(A_0, b|m)} = \begin{cases}
        s_{(j)}^{A_0} \bigg( 1 + \frac{\sum_{i=m+1}^{n} s^{A_0}_{(i)}}{\sum_{i=1}^{m} s^{A_0}_{(i)}} \bigg)  & j \le m \\
        0 & j > m
        \end{cases} 
        $$

        $$x_{(j)}^{F(A_0, b|m)} = \begin{cases}
        -s_{(j)}^{A_0} \frac{\sum_{i=m+1}^{n} s^{A_0}_{(i)}}{\sum_{i=1}^{m} s^{A_0}_{(i)}} p  & j \le m \\
        s_{(j)}^{A_0} p & j > m
        \end{cases}$$
\end{definition}

The (expected) utility of agent $i$ under the bidding profile $b=(b_j)_{j \in I}$ and initial allocation $A_0$ is then given by:
    $$
    u^F_i(A_0,b) = P(m=\bar{m}) \cdot u_i(F(A_0,b|\bar{m})) + P(m=\bar{m}-1) \cdot u_i(F(A_0,b|\bar{m}-1))
    $$
where we usually omit the subscript $F$ when clear from context.
Below we introduce the mechanism characteristics we wish to analyze. These properties are motivated by those studied in \cite{myerson1983efficient} for the bilateral trading framework.
We start by defining the well-known notion of a ``strategyproof" mechanism. A mechanism is considered \textit{strategyproof} if bidding one's private value is a weakly dominant strategy for all agents, regardless of all the other agent bids. Formally:
	\begin{definition}
		A mechanism is considered \textit{strategyproof} if for every agent $i \in I$, every $(b_j)_{j \in I} \in \mathbb{R}^n$, and every $A_0 \in \mathcal{A}$, it holds that
        $$u_i(A_0,(\bar{b}_i,b_{-i})) \geq u_i(A_0,b)$$
        
		where $b_{-i}$ are the bids of all the agents other than $i$.
	\end{definition}
	
	While the definition above tackles the case in which a single agent deviates, our mechanism also resolves cases in which a subset of agents all benefit from working in collusion. Next, we define the notion of a weak group strategyproof mechanism. A mechanism is considered \textit{weak group strategyproof} if it is impossible for all the agents who deviate from bidding their private valuation to strictly benefit simultaneously. Formally:
	
	\begin{definition}
		A mechanism is considered \textit{weak group strategyproof} if there does not exist a subset of agents $J \subseteq I$, bids $(b_i)_{i \in J} \in \mathbb{R}^{|J|}$, and initial allocation $A_0 \in \mathcal{A}$, such that for every agent $j \in J$, it holds that

        $$u_j\left(A_0,((b_i)_{i \in J},(\bar{b}_i)_{i \in I \setminus J}))\right) > u_j(A_0,\bar{b})$$
  
	\end{definition}
	
	Next, we define the notion of a budget balanced mechanism. Note that our definition of budget balanced, also includes a requirement that the asset will be fully allocated.
	\begin{definition}
		We say that a mechanism $F$ is \textit{budget balanced} if for every $b=(b_i)_{i \in I} \in \mathbb{R}^n$, every initial allocation $A_0 \in \mathcal{A}$, and every allocation $A \in \mathcal{A}$ in the support of $F(A_0,b)$, it holds that
		$$\sum_{i \in I} s^{A}_i = \sum_{i \in I} s^{A_o}_i$$
		and
		$$\sum_{i \in I} x^{A}_i = \sum_{i \in I} x^{A_o}_i$$
	\end{definition}

        When considering a strategyproof mechanism, it is sufficient to discuss the outcome of the mechanism in the case where agents play their dominant strategy, which is bidding their true valuations. Hence the following two properties, which are related to the quality of the outcome, are only defined with respect to truthful bidding of the agents.
        
         While it is clear that efficiency can only be obtained by allocation of the asset to the agent with the highest valuation, many real-life applications require distributing the asset among a larger, predefined number of agents. A natural weaker requirement of such distribution is that the asset is allocated to the set agents who value it the most. While there are infinitely many such allocations, a second natural requirement is that the allocation preserves the proportional share of each agent compared to the initial allocation. That is, an allocation $A$ is said to be proportionality-preserving with respect to initial allocation $A_0$ if, for every pair of owners in $A$, the ratio between their eventual shares is the same as the share between their initial shares. \textit{Proportionality-preserving Ex-post efficiency} is then defined as follows:

        \begin{definition}
            We say that a mechanism $F$ satisfies \textit{proportionality-preserving ex-post efficiency} if for every $A_0 \in \mathcal{A}$, 
            any allocation $\bar{A} \in \mathcal{A}$ in the support of $F(A_0,\bar{b})$ satisfies the following two conditions:
            \begin{enumerate}
                \item For any two agents $j, k \in I$ such that $s_j^{\bar{A}} > 0$ and $s_k^{\bar{A}} > 0$ (both agents are buyers), it holds that $\frac{s_j^{\bar{A}}}{s_k^{\bar{A}}} = \frac{s_j^{A_0}}{s_k^{A_0}}$.
                \item There exists no pair of agents $j, k \in I$ such that $s_j^{\bar{A}} = 0$ (agent $j$ is a seller), $s_k^{\bar{A}} > 0$ (agent  $k$ is a buyer), and $v_j > v_k$. 
            \end{enumerate}
        \end{definition}

        Finally, we define \textit{individual rationality}, which implies that under truthful bidding all agents improve their utility with respect to the initial allocation:
        
	\begin{definition}
		We say that a mechanism $F$ satisfies \textit{individual rationality} if for every $i\in I$ and every $A_0 \in \mathcal{A}$, it holds that
        $$u_i(A_0,\bar{b}) \ge u_i(A_0)$$
	\end{definition}
	
	Lastly, we define the useful notion of \textit{adjusted utility}, which is simply the difference between the agent's utility before and after applying a restructuring mechanism:
 \begin{definition}
     Given an allocation $A \in \mathcal{A}$, the \textit{adjusted utility} of agent $j$ with respect to the initial allocation $A_0 \in \mathcal{A}$ is defined as follows:
     $$\tilde{u}_j(A) = u_j(A) - u_j(A_0)$$
     Similarly, the adjusted utility under a mechanism $F$ and a bidding profile $b=(b_i)_{i \in I}$ is:
     $$\tilde{u}_j(A_0,b) = u_j(A_0,b) - u_j(A_0)$$
 \end{definition}

 Note that individual rationality can be redefined by requiring that the adjusted utility of each agent with respect to the allocation obtained by truthful bidding is non-negative. Moreover, note that since the initial utilities are constant, properties like strategyproof and weak group strategyproof can be equivalently defined with respect to the adjusted utility, as agents always have a beneficial deviation with respect to their standard utility if and only if they have a  beneficial deviation with respect their adjusted utility. Therefore, from now on we only consider the agents' adjusted utility, without explicitly specifying it.
	

	
\section{Analysis and Results}
In this section we show that the MBM is strategyproof, weakly group strategyproof, budget balanced, individually rational and proportionality-preserving ex-post efficient. An essential part of our mechanism is the monotone relation between the bids and the price. The next lemma shows that if an agent increases her bid, then the mechanism sets a weakly higher price. If she decreases her bid then the mechanism sets a weakly lower price.
	\begin{lemma}	
      \label{lemma - change one bid}
		let $b,b^h,b^l$ be bid vectors such that $b^h$ is derived by increasing one bid in $b$ and $b^l$ is derived by decreasing one bid in $b$. Let $p, p^h, p^l$ be the mechanism prices corresponding to the vector bids of $b,b^h,b^l$, respectively. 
		Then, $p^l \leq p \leq p^h$.
	\end{lemma}
	
	\proof{}
		Denote by $h,l$ the agents who bids higher and lower in $b^h, b^l$, respectively. 
		By increasing agent $h$'s bid, the value associated with each ranking weakly increases, and in particular the value associated with the $m$'th ranking weakly increases from $p$ to $p^h$. Similarly, by decreasing agent $l$'s bid, the value associated with the $m$'th ranking weakly decreases from $p$ to $p^l$.
	\endproof
	
	To show strategyproofness we take advantage of a characteristic for each type the mechanism can assign to agent $j:$
	a \emph{definite buyer} (i.e., $r(j) < \bar{m}$),
	a \emph{definite seller} (i.e., $r(j) > \bar{m}$),
	and \emph{the threshold agent} (i.e., $r(j) = \bar{m}$).
	If agent $j$ is a definite buyer then her utility equals $~{s^{A_0}_{j}\frac{\sum_{i=m+1}^{n} s^{A_0}_{(i)}}{\sum_{i=1}^{m} s^{A_0}_{(i)}}(v_j - p)}$, which is positive if and only if the asset price is lower than she values it. 
	If the mechanism designates agent $j$ as a definite seller then her utility is $s^{A_0}_{j}(p - v_j)$, which is positive whenever the asset price is higher than her valuation of it.  
	Finally, if $r(j)=\bar{m}$, then the agent can be designated both as a buyer and as a seller. We observe that according to the probabilities for each realization, the agent's expected utility is zero.
	\begin{observation}	\label{obs: zero utility}
		The threshold agent's expected utility is zero.
	\end{observation}
	
	\proof{}
		Let $j \in I$ such that $r(j)=\bar{m}$. Agent $j$ can be either a buyer or a seller, but the price will always be $p = b_{(\bar{m})}$. Therefore, it holds that
		$$\sum_{i=1}^{\bar{m}} s^{A_0}_{(i)} \left[s^{A_0}_{j}\frac{\sum_{i=\bar{m}+1}^{n} s^{A_0}_{(i)}}{\sum_{i=1}^{\bar{m}} s^{A_0}_{(i)}}(v_j - p)\right] + \sum_{i=\bar{m}+1}^{n} s^{A_0}_{(i)} \left[s^{A_0}_{j}(p - v_j)\right] = 0$$
	\endproof
	
	We also observe that since the mechanism sets the $\bar{m}$'th bid as the price, under $\bar{r}$ all agents receive a non-negative utility, which implies that the mechanism satisfies individual rationality:
	\begin{lemma}\label{lem:positive_util}
		The MBM satisfies individual rationality.
	\end{lemma}
	
	\proof{}
		Recall that under $\bar{r}$ every agent $i \in I$ bid her private valuation $\bar{b}_i = v_i$,
		and the price is $\bar{p} = \bar{b}_{(\bar{m})}$.
		In addition, for every definite buyer $i$, since $r(i) < \bar{m}$, it holds that $\bar{b}_i \geq \bar{b}_{(\bar{m})}$. Therefore, $v_i \geq \bar{p}$. Since the difference $v_i - \bar{p} \geq 0$ the buyer utility is non-negative.
		Similarly, for every definite seller the difference $\bar{p} - v_i \geq 0$, thus the seller utility is non-negative as well.
	\endproof
	
	Next, we show that agents will play their type in equilibrium.
	
	\begin{lemma}	\label{lem: strategyproofness}
		The MBM is strategyproof.
	\end{lemma}
	
	\proof{}
    Let $i \in I$, and let $b_{-i}$ be some arbitrary bidding profile of all players except player i (not necessarily truthful bids).
        We show that player $i$ is weakly prefers bidding $b_i=v_i$:

        First, assume that bidding truthfully makes agent $i$ a definite buyer. In this case, the price must be weakly lower than her valuations (otherwise she was a seller). Increasing her bid does not change her utility. Decreasing her bid will only change her utility when she hits the price $p$. If she bids the price $p$ she becomes the threshold agent and gets utility $0$, and if she gets lower than the price then she becomes a seller, but she sells at a price lower than her valuations - which leads to a negative utility, and therefore this is not beneficial for her. Similar arguments can be applied to the case where bidding truthfully makes agent $i$ a definite seller. 
        
        Lastly, if bidding truthfully makes agent $i$ a threshold agent, her expected utility is $0$, and it is straightforward to see that either increasing or decreasing her bid is not beneficial: increasing the bid either leaves her as the threshold agent (and then she obtains zero payoff, regardless of the price), or make her a definite buyer - but with a price higher than her valuation which causes negative payoff. Similarly, decreasing the bid either leaves her as the threshold agent, of causes a negative utility due to selling at a price lower than her valuation.
	\endproof
	
    Proportionality-preserving ex-post efficiency of the MBM then follows as an immediate corollary, as when agents bid their true valuations, the highest bidders remain owners (and proportionality of buyer's shares is trivially preserved by definition of the mechanism):

    \begin{corollary}
    \label{cor:ex post pareto}
    The MBM is proportionality-preserving ex-post efficient.
    \end{corollary}
 
 By Lemma \ref{lem: strategyproofness}, every agent is better off bidding her true valuation. However, if agent $j$ is ranked $m$'th and $b_{(m-1)} < b_{(m)} < b_{(m + 1)}$, she increases sellers' (buyers') payoff by increasing (decreasing) her bid and the price by a small amount, while maintaining her ranking. Thus, the MBM is not strong group strategyproof. 
	However, to show that the mechanism is weakly group strategyproof we need to show that it is impossible for all the agents who deviate from bidding their true valuation to strictly benefit simultaneously. We show this in the next lemma:
	
	\begin{lemma}	\label{lem: weak group strategyproofness}
		The MBM is weakly group strategyproof.	
	\end{lemma}
	
	\proof{}
		Assume contrary to the lemma that there exists a set of agents $J \subseteq I$ such that $|J| \geq 2$ that do not bid their private valuation under $r$, and that all agents in $J$ strictly gain by deviating from $\bar{r}$ to $r$.
		
		By Lemma \ref{lem:positive_util}, all agents get non-negative utility under the truthful bidding profile. Let $J_B$, $J_S \subseteq J$ the definite buyers and sellers in $J$ under $r$, respectively. $J = J_B \cup J_S$ (i.e., no deviating agent is the threshold agent under $r$). To see why, suppose that there exists agent $j \in (J \setminus J_B) \setminus J_S$, then $r(j) = \bar{m}$, and her expected payoff is zero. Denote by $\bar{J_B}, \bar{J_S} \subseteq J$ the definite buyers and sellers in $J$ under $\bar{r}$, respectively, and denote by $\bar{j}$ the threshold agent for which $\bar{r}(\bar{j}) = \bar{m}$.
		
        Because the subset size is fixed by the mechanism, if $J_B \subseteq \bar{J_B}$ then $J_B = \bar{J_B}$. Therefore, $J_S = J \setminus J_B = J \setminus \bar{J_B} \supseteq (J \setminus \bar{J_B}) \setminus \{\bar{j}\} = \bar{J_S}$, and because the set size is fixed it holds that $J_S = \bar{J_S}$. Namely, in this case, all the definite buyers remain definite buyers and all the definite seller remain definite sellers after subset $J$ of agents deviate from $\bar{r}$. Thus, the price and shares in $r$ remain as in $\bar{r}$, and no agent gains by deviating. Similarly, no agent gains if $J_S \subseteq \bar{J_S}$.
	
        Hence, if the deviating agents can simultaneously strictly gain, then there exist $j_b \in J_B \cap \{J \setminus \bar{J_B}\}$ and $j_s \in J_S \cap \{J \setminus \bar{J_S}\}$. That is, $r$ must produce a new definite buyer and a new definite seller. Since agents bid their private valuation under $\bar{r}$, it holds that $v_{j_b} \leq \bar{p} \leq v_{j_s}$. Therefore, for every price $p$ either $v_{j_b} - p \leq 0$ or $p - v_{j_s} \leq 0$. Thus, at least one agent does not strictly increase her expected utility.
	\endproof
	
	The mechanism is designed such that all the seller shares are distributed to the buyers. In addition, the price is tied to the asset itself, and the share price is the asset price multiplied by share size. The next lemma shows that the MBM is budget balanced.
	\begin{lemma}	\label{lem: budget balance}
		The MBM is budget balanced.
	\end{lemma}
	\proof{}
		Recall that $A_0$ is the initial allocation. The summation of the realized buyers' additional shares is equal to the summation of the realized sellers' initial shares:
		$$ \sum_{j=1}^{m} s^{A_0}_{(j)} \frac{\sum_{i=m+1}^{n} s^{A_0}_{(i)}}{\sum_{i=1}^{m} s^{A_0}_{(i)}}
		= \frac{\sum_{i=m+1}^{n} s^{A_0}_{(i)}}{\sum_{i=1}^{m} s^{A_0}_{(i)}} \sum_{j=1}^{m} s^{A_0}_{(j)}
		= \sum_{i=m+1}^{n} s^{A_0}_{(i)} $$
		In addition, because a price $p$ is set for the asset itself, combined with the previous equation, the summation of payments and compensations over all agents yields:

        $$\sum_{j=1}^{m} s_{(j)}^{A_0} \frac{\sum_{i=m+1}^{n} s^{A_0}_{(i)}}{\sum_{i=1}^{m} s^{A_0}_{(i)}} p - \sum_{j=m+1}^{n} s_{(j)}^{A_0} p = 0$$

		and the sum of agents' initial money is also zero. Thus, the mechanism is balanced in both shares and money.
	\endproof
	
	We derive the main theorem directly by combining lemmas \ref{lem:positive_util}, \ref{lem: strategyproofness}, \ref{lem: weak group strategyproofness}, \ref{lem: budget balance} and corollary \ref{cor:ex post pareto}:
	
	\begin{theorem}
		The MBM is strategyproof, weakly group strategyproof, budget balanced, individually rational and proportionality-preserving ex-post efficient.
	\end{theorem}

\subsection{Welfare Considerations}
\label{subsec: welfare}

Social welfare among the agents is an important aspect of a restructuring mechanism. While traditionally, mechanisms are often designated to maximize social welfare or the seller's revenue, our MBM aims at balancing agents' welfare and two other objectives of the designer: the eventual number if owners (which are endogenously set by the designer), and preservation of the shares proportionality among shareholders post-restructuring.

Although MBM is not explicitly designed to maximize social welfare, it nevertheless improves upon the initial allocation. To understand this improvement, we can conceptualize the social welfare using probability and expected value frameworks.

Initially, we view the allocation as a probability distribution over agents' valuations, where each valuation $v_i$ is associated with probability $s^{A_0}_i$. The initial social welfare is thus the expectation $\mathbb{E}[v]$.
The MBM introduces a random variable $m$, taking values of either $\bar{m}$ or $\bar{m}-1$, which represents the eventual number of owners. This randomness is crucial for the mechanism's strategyproofness. For a given $m$, the MBM allocates the entire asset to the $m$ highest-valuing agents, preserving their initial share proportionality.
The resulting social welfare can be understood as a weighted sum of valuations. With probability $P(m=\bar{m})$, it is the sum of the top $\bar{m}$ agents' valuations, and with probability $P(m=\bar{m}-1)$, it is the sum of the top $\bar{m}-1$ agents' valuations. In both cases, each valuation is weighted by the agent's proportional share among remaining owners.

This method ensures that the expected social welfare after applying the MBM is always at least as high as the initial welfare. The improvement stems from concentrating ownership among those who value the asset most highly, while preserving proportionality among remaining owners.  
While the optimal allocation (giving the whole firm to the highest bidder) is never achieved due to proportionality preservation, the MBM still improves welfare.
Importantly, the total welfare post-MBM is solely the sum of the remaining owners' valuations. Compensation to sellers is an internal transfer and doesn't affect the overall welfare calculation.

The welfare improvement can be understood by considering extreme cases. When $\bar{m}=n$, social welfare remains unchanged (as $m=\bar{m}=n$ with probability 1). As $\bar{m}$ decreases, the expected social welfare increases. This relationship provides a method for choosing $\bar{m}$: designers can maximize the number of remaining shareholders while maintaining a desired level of social welfare.
\footnote{One can view social welfare in terms of conditional expectation. Let $\mu$ be the probability distribution of agents' valuations from the initial allocation. The initial social welfare is $\mathbb{E}_{\mu} [v]$. For a realization of $m$, the MBM allocation's social welfare is the conditional expectation of $v$ being at least the $m$'th highest value, due to preserved proportionality.}
To evaluate the mechanism's effectiveness, we can consider the efficiency loss—the ratio between the expected social welfare under MBM and the welfare under the first-best allocation (where the entire asset goes to the highest-valuing agent). In general, this ratio can be arbitrarily small. For instance, if the highest-valuing agent has an arbitrarily small initial share, their post-restructuring share (and thus the social welfare) can be arbitrarily small for any $\bar{m} < n$.
However, in the more natural case of equal initial shares and uniform agent valuations, the efficiency loss is bounded, especially with many agents. In this scenario, both social welfare and efficiency loss decay linearly with the fraction of remaining owners, $\frac{\bar{m}}{n}$. Further details on this analysis can be found in Appendix \ref{app: equal shares}.

\section{Discussion}
In this work we present an intuitive mechanism for ownership restructuring in private firms. 
	Our newly introduced mechanism reallocates shares of a shared asset with the goal of decreasing the number of owners as needed.
	The mechanism is robust to strategic behavior and collusion (weak group strategyproofness), and maintains budget balance. While \citet{Cramton1987} achieve ex-post efficiency for some initial shares distributions, our mechanism satisfies a weaker notion of proportionality-preserving ex-post efficiency, regardless of the initial allocation.
	
	The mechanism receives bids from the original owners. Based on the bids and the original allocation, it decides on a new allocation of shares and a single price that is used for all the transactions.
	In the resulting allocation, the remaining owners increase their shares in proportion to their original holdings in exchange for payments that are lower than their valuation of their additional share. The rest of the original owners are compensated with payments that are higher than their valuation for the share they forfeit.
	
	An alternative mechanism that generalizes the BMBY rationale in order to solve a multi-agent scenario is a series of 2-agent BMBY ``matches'', where the share holder who wins a match and remains an owner is matched with the next share holder, until the number of owners decreases as required. Yet another alternative is having multiple 2-agent BMBY run simultaneously, and rematch the remaining owners with each other, until the required number of owners is reached. In both mechanisms the outcome depends heavily on the agents that are selected for a match and the order of matches. Hence, such mechanisms could let low value agents remain owners while high value agents sell their share. In addition, these alternatives are not strategyproof since an agent could bid high to win one match only to sell at a higher price in the next match. 
	
	In contrast to the above, the order of bids in the MBM does not affect the outcome, and it can be implemented when agents observe prior bids. Besides efficiency, budget balance, and weak group strategyproofness, a notable strength of our mechanism is its low computational time complexity. Moreover, the mechanism is easy to comprehend and to implement, which makes it a fitting business tool for ownership restructuring.

    Importantly, the MBM permits a dynamic implementation via an ascending clock auction, which endows the agents with obviously dominant strategies, implying that the MBM is \emph{obviously strategyproof} (\citealp{li2017obviously}). Such an implementation enhances its reliability, elevates trust and participation, and indicates robustness to agents' bounded rationality. Interestingly, this approach also provides an alternative proof of the mechanism's weak group strategyproofness.

    Lastly, according to Myerson’s revelation principle, it is possible to construct a direct mechanism that implements truthful revelation from any given restructuring mechanism. However, this approach does not guarantee any clear characterization of the resulting mechanism, and many of its economic properties remain uncertain. Our proposed mechanism turns out to have a simple and intuitive construction, that improves upon the existing BMBY mechanism by eliciting agents’ valuations, being applicable beyond the two-agent case, and providing a clear welfare analysis.
	
	An interesting follow-up work may include the development of similar mechanisms for the case of non-linear utilities, or adding additional constraints such as upper-bounding agents' shares. For instance, one may require that no agent holds more than 50\% of the asset after the restructuring.
    Additionally, note that there is a gap between the willingness to pay of the highest ranking seller and the valuation of the lowest ranking buyer. One may consider devising a mechanism in which the firm utilizes this gap to raise funds, as is often the reason for ownership restructuring. 
	


\bibliographystyle{plainnat}
\bibliography{references}

\appendix
\section{Welfare Analysis: Equal Shares and Uniform Valuations}
\label{app: equal shares}

Consider the case of $n$ agents, whose valuations are given by $v_i = \frac{n - i + 1}{n}$ for $i = 1, \dots, n$, and each agent has an initial share of $\frac{1}{n}$. The designer is willing to cut ownership by a constant factor of $1-\alpha$, meaning she applied the MBM with $\bar{m} = \alpha n$, where $\alpha \in \left\{\frac{2}{n}, \dots, \frac{n-1}{n}\right\}$. The eventual owners will be either the first $\bar{m}$ agents with probability $\frac{\bar{m}}{n}=\alpha$, or the first $\bar{m}-1$ agents with the complementary probability. Therefore, the expected social welfare under the MBM is given by:

\[
SW = \alpha \frac{1}{\bar{m}} \sum_{i=1}^{\bar{m}} v_i + (1-\alpha) \frac{1}{\bar{m}-1} \sum_{i=1}^{\bar{m}-1} v_i
\]

and the first best welfare is simply $v_1 = 1$. Hence, $SW$ is also the fraction of welfare preserved by the MBM.\footnote{By \cite{Cramton1987}, the first-best allocation is indeed achievable by a strategyproof, budget-balanced, and individually rational mechanism in the case of equal shares (for any number of agents), hence it is the correct benchmark in this case.} We show that as the number of agents grows large and $\alpha$ remains constant, this fraction cannot get below $\frac{1}{2}$, meaning that the MBM preserves at least $50\%$ of the social welfare. Additionally, the efficiency loss decays linearly in $\alpha$.

We first simplify the sums of $v_i$. The sum of the first $\bar{m}$ terms of $v_i$ is:

\[
\sum_{i=1}^{\bar{m}} v_i = \frac{\bar{m}}{n} \cdot \frac{2n - \bar{m} + 1}{2}
\]

Similarly, for the first $\bar{m} - 1$ terms:

\[
\sum_{i=1}^{\bar{m}-1} v_i = \frac{\bar{m} - 1}{n} \cdot \frac{2n - \bar{m} + 2}{2}
\]

We substitute these sums into the expression for the social welfare:

\[
SW = \alpha \cdot \frac{1}{n} \cdot \frac{2n - \bar{m} + 1}{2} + (1 - \alpha) \cdot \frac{1}{n} \cdot \frac{2n - \bar{m} + 2}{2}
\]

Factoring out common terms and simplifying:

\[
SW = \frac{1}{2n} \bigg( \alpha (2n - \bar{m} + 1) + (1 - \alpha)(2n - \bar{m} + 2) \bigg) = \frac{1}{2n} \left( 2n - \bar{m} + 2 - \alpha \right)
\]

Substitute $\bar{m} = \alpha n$ and further simplify:

\[
SW = \frac{(2 - \alpha)n + 2 - \alpha}{2n} = \frac{2 - \alpha}{2} + \frac{2 - \alpha}{2n}
\]

As $n \to \infty$, the second term vanishes, so:

\[
\lim_{n \to \infty} SW = \frac{2 - \alpha}{2} \in \left( \frac{1}{2}, 1 \right)
\]

\end{document}